\documentstyle[11pt,aaspp4,tighten]{article}

\def\today{\rightline{\ifcase\month\or
        January\or February\or March\or April\or May\or June\or
        July\or August\or September\or October\or November\or December\fi
        \space\number\day, \number\year}}
\def\etal{{\it et al.}\ }

\def\s-1{{\rm\,s^{-1}}}

\def\spose#1{\hbox to 0pt{#1\hss}}
\def\C3H2{{\rm\,\rm C_3H_2}}
\def\NH3{{\rm\,\rm NH_3}}
\def\HOCO+{{\rm\,\rm HOCO^+}}
\def\lta{\mathrel{\spose{\lower 3pt\hbox{$\mathchar"218$}}
     \raise 2.0pt\hbox{$\mathchar"13C$}}}
\def\gta{\mathrel{\spose{\lower 3pt\hbox{$\mathchar"218$}}
     \raise 2.0pt\hbox{$\mathchar"13E$}}}

\begin{document}

\font\twelvei = cmmi10 scaled\magstep1 
       \font\teni = cmmi10 \font\seveni = cmmi7
\font\mbf = cmmib10 scaled\magstep1
       \font\mbfs = cmmib10 \font\mbfss = cmmib10 scaled 833
\font\msybf = cmbsy10 scaled\magstep1
       \font\msybfs = cmbsy10 \font\msybfss = cmbsy10 scaled 833
\textfont1 = \twelvei
       \scriptfont1 = \twelvei \scriptscriptfont1 = \teni
       \def\mit{\fam1 }
\textfont9 = \mbf
       \scriptfont9 = \mbfs \scriptscriptfont9 = \mbfss
       \def\bmit{\fam9 }
\textfont10 = \msybf
       \scriptfont10 = \msybfs \scriptscriptfont10 = \msybfss
       \def\bmsy{\fam10 }

\def\etal{{\it et al.~}}
\def\eg{{\it e.g.}}
\def\ie{{\it i.e.}}
\def\lsim{\raise0.3ex\hbox{$<$}\kern-0.75em{\lower0.65ex\hbox{$\sim$}}} 
\def\gsim{\raise0.3ex\hbox{$>$}\kern-0.75em{\lower0.65ex\hbox{$\sim$}}} 
\title{A SURVEY FOR INFALL MOTIONS TOWARD STARLESS CORES.\\
I. $\rm CS~(2-1)$ AND $\rm N_2H^+~(1-0)$ OBSERVATIONS}

\author{Chang Won Lee$^{1,2}$, Philip C. Myers$^{1}$, \& Mario Tafalla$^{1,3}$}
\vskip 0.2in
\affil{$^1$Harvard-Smithsonian Center for Astrophysics,}
\affil {60 Garden Street, MS 42, Cambridge, MA  02138, USA}

\vskip 0.2in
\affil{$^2$Taeduk Radio Astronomy Observatory, Korea Astronomy Observatory,}
\affil {36-1 Hwaam Dong, Yusung Ku, Taejon 305-348, Korea}
 
\vskip 0.2in
\affil{$^3$Observatorio Astron\'omico Nacional,}
\affil {Campus Universitario, Apartado 1143, E-28800, Alcal\'a de Henares 
(Madrid), Spain}
\affil{E-mail: cwl@hae.issa.re.kr, pmyers@cfa.harvard.edu, tafalla@oan.es}

\vskip 1in
\begin{abstract}
We present the first results of a survey of 220 starless cores selected 
primarily by their optical obscuration (Lee \& Myers 1999) and observed
in CS(2--1), N$_2$H$^+$ (1--0), and C$^{18}$O (1--0) 
using the NEROC Haystack 37-m telescope. 
We have detected 163 out of 196 sources observed in CS, 72 out of 142 in N$_2$H$^+$, 
and  30 out of 30 in C$^{18}$O.
In total, 69 sources were detected in both CS and N$_2$H$^+$.

The isolated  component of the N$_2$H$^+$ (1--0) spectrum ($\rm F_1F=01-12$)
usually shows a weak symmetric profile which is 
optically thin. In contrast, a
significant fraction of the CS spectra show non-Gaussian shapes, which
we interpret as arising from a combination
of self absorption due to lower excitation gas in the core front
and kinematics in the core.

The distribution of the normalized velocity difference ($\delta V_{CS}$) 
between the CS and N$_2$H$^+$ peaks appears significantly skewed to the blue 
($\delta V_{CS} < 0$), 
as was found in a similar study of dense cores with embedded
young stellar objects (YSOs) (Mardones et al. 1997). 
The incidence of sources with blue
asymmetry tends to increase as the total optical depth or 
the integrated intensity of the N$_2$H$^+$ line increases. 
This overabundance of ``blue'' sources  over ``red'' sources
suggests that inward motions are a significant feature of 
starless cores.  
We identify 7 strong infall candidates and 10 probable infall candidates.
Their typical inward speeds 
are sub-sonic, approximately $\rm 0.04 - 0.1~km~s^{-1}$, so they 
contain `thermal' infall motions, unlike the faster inward 
speeds associated with most YSOs (Mardones et al. 1997).
We discuss the importance of the choice of a consistent set
of line frequencies when using the velocity shift between an 
optically thick and a thin line as a tracer of infall, and
show how the results of the survey depend on that frequency choice.
\end{abstract}


\clearpage

\section{Introduction}

A ``starless'' core is a dense core with no embedded IRAS source 
and no associated T Tauri star, implying that
it is in the earliest identifiable stage of star formation (Shu, Adams,
\& Lizano 1987; Ward-Thompson et al. 1994; Lee \& Myers 1999, hereafter LM99).
Low mass stars are believed to form through the gravitational collapse of 
such a core when its central density becomes high enough (Shu 1977).

The last decade has seen remarkable progress on the understanding of 
low mass star formation.
However, most studies of infall phenomena have been focussed 
on Young Stellar Objects (YSOs)
or cores with embedded YSOs (Zhou et al. 1993; Gregersen et al. 
1997; Mardones et al. 1997 
hereafter M97).
Most of our understanding of the physics of the pre-stellar stage  
is still entirely theoretical
(e.g., Shu, Adams, \& Lizano 1987, Ciolek \& Mouschovias 1995). 
A systematic observation of the starless cores has been 
carried out in continuum (Ward-Thompson et al. 1994, 1999), but
not yet in spectral lines sensitive to inward motions.
Thorough observational studies of starless cores in molecular 
lines are required to 
understand initial conditions for star formation and evolution
of star-forming cores.

The physical and kinematical state of starless cores is still unclear.
Using sub-millimeter data of 21 starless cores, Ward-Thompson et al. (1994) 
suggested  that starless cores are currently in the  ambipolar 
diffusion phase prior to protostellar collapse. 
However, combining observations of  optically thick and thin lines,
Tafalla et al. (1998) and Williams et al. (1999)
have shown that even a starless core like L1544 can present many 
symptoms of advanced evolution toward star formation.  
These authors found that the large area with infall asymmetry in 
L1544 is inconsistent  with 
the `inside-out' collapse model (Shu 1977), and that
the large inward speed (up to 0.1 km s$^{-1}$)
is also too fast to result from ambipolar diffusion.
Therefore L1544 was interpreted as having some type of inward motions
in a manner not explained by any present `standard' star formation 
model. 
We still do not know whether L1544 is an exceptional case
or it represents a phase that every core experiences 
in the early course of star formation.

Systematic and extensive observational studies of starless cores may allow us 
to find good answers for the above questions.
Recently, Lee \& Myers (1999) have presented a new catalogue of 
dense cores optically selected
by using the STScI Digitized Sky Survey,  where 306 starless cores 
are identified.
We have made extensive multi-molecular line surveys toward starless cores,
mainly based on the list of starless cores in the LM99 catalogue which 
provides the largest available database to study starless cores 
with a statistical approach.
Here we report the first result of one such survey
using optically thick [CS (2--1)] and thin 
[N$_2$H$^+$ (1--0) and C$^{18}$O (1--0)] tracers, as  the first step 
to study the physical and kinematical status of starless cores.
Our main goals in this paper are to quantify the 
incidence of infall motions toward starless cores and 
to find infall candidates for further study in the future. 
 
In $\S$2 we explain how we selected and observed the starless cores, 
and reduced the molecular line data.
Results of the qualitative and quantitative analysis of 
the molecular line survey data are 
presented in $\S$3. Implications of our statistics on starless
cores, identification of infall candidates, 
kinematics of the suggested infall candidates, and future work
are discussed in $\S$4; we summarize our results in $\S$5.

\section{Observations}
We have conducted single-pointing observations for 220 starless cores
in the lines of CS (2--1), N$_2$H$^+$ (1--0), and C$^{18}$O (1--0).
Most source positions were chosen from 
the catalogue of optically selected dense cores by LM99, to which
we added 3 new positions (L1622A-2, L183B, and L1155C-2) and slightly 
modified 17 within about $1' - 2'$ using
our recent FCRAO maps (Lee, Myers, \& Tafalla 1999) so that they coincide
with molecular peaks. 
Nine of the 220 positions were adopted from Benson \& Myers (1989), 
although 7 of them are listed in LM99 catalogue 
with similar coordinates because the LM99 catalogue was not available 
at the beginning of our survey.
Tables 1 and 2 list the names and positions of the observed sources,
together with the position references.
Including the observations of all sources with similar positions to those
in the LM99 catalogue, we have observed about 93$\%$ (212) of the 227 
sources in the catalogue which are observable from Haystack 
($\delta \ga -25^\circ$). 

In order to understand the optical depth
and kinematical characteristics of the spectra, 
it is necessary to observe both optically thick and thin molecular lines.
In this way, we can distinguish profile asymmetries that arise from 
self absorption from those due to multiple components in the
emitting gas.
We have chosen CS (2--1) as our thick tracer and N$_2$H$^+$ (1--0) as our
thin tracer because they have been already successfully used in identifying
infall signatures in YSOs (e.g., M97). 
The  C$^{34}$S isotope would be the ideal thin partner for CS, as 
it probes the same physical properties (temperature,
density, chemistry, and size scale) as the main isotope. However,
it is usually too weak to determine its peak velocity with a good 
signal-to-noise (S/N) ratio, and for this reason, we have chosen 
N$_2$H$^+$. The N$_2$H$^+$ (1--0) line at 93 GHz consists
of 7 hyperfine components (Caselli, Myers, \& Thaddeus 1995; see also Fig. 5), 
allowing us to determine the peak velocity
with about twice the precision of a single line with the same 
S/N ratio.  The isolated component ($\rm F_1F=01-12$) 
is usually optically thin (M97), and does not show self-absorption 
features for most sources. The C$^{18}$O (1--0) line was used as a 
substitute optically thin tracer for several sources not detected in 
N$_2$H$^+$ (1--0).

For the quantitative study of spectral line asymmetries, it is necessary to 
use accurate and consistent frequencies for the different molecular lines.
Despite the high accuracy of laboratory microwave spectroscopy,
the frequency uncertainties in standard line catalogs (e.g, Lovas 1992, 
Pickett et al. 1998) are of the order of tens of kHz, and therefore close to 
some of the observed line velocity differences (few tenths of 
km s$^{-1}$, see Section 4). This
makes the choice of the line frequencies potentially critical 
in our statistical study of velocity differences between the peaks of
the thick and the thin tracers, and a careful consideration of the frequency
values is necessary. For this reason we have adopted the 
frequencies recommended by Tafalla et al. (1998), which seem to 
constitute a consistent set:
97980.950 MHz for CS (2--1) from Pickett et al. (1998),
109782.182 MHz  for C$^{18}$O (1--0) from Lovas \& Krupenie (1974), and  
93176.2651 MHz for the $\rm N_2H^+~(1-0)$ 
``isolated'' component ($\rm F_1F=01-12$) from  Caselli et al. (1995).
In the Appendix, we describe in detail how and why this set 
of frequencies was chosen.

Our observations were made using the NEROC Haystack 37m telescope\footnote[1]
{Radio astronomy observations at 
the Haystack Observatory of the Northeast Radio Observatory Corporation are 
supported by a grant from National Science Foundation}
in 1997 February, March, and December, and in 1998 January, March, and May.
Two receivers were used simultaneously to obtain spectra in the
left and right circular polarizations, and the signals were later combined.
The back-end was an autocorrelator with 17.8 MHz bandwidth and 
2.17 kHz (0.007 km s$^{-1}$) spectral resolution.
The spectra were obtained in frequency-switching mode with a frequency
throw of $\pm$ a quarter of the bandwidth -- $\pm$4.425 MHz.
The full width at half maximum (FWHM) of the telescope beam is about 
$27''$ at 86 GHz  (Barvainis et al. 1993), and the telescope main beam 
efficiencies are $18\pm3\%$ at 86 GHz and 
$16.9\pm2\%$ at 115 GHz (Ball et al. 1993).  
The total integration time for each spectrum was usually between 
30 and 120 minutes, depending on the source brightness, in order to achieve
a S/N ratio better than $\sim 8$.
Pointing was checked by using mostly Venus, 
Saturn, and Jupiter whenever the region of the observing source was changed.  
The usual pointing error was within about $6''$.

\section{Results}
\subsection{Detection Statistics}

In total, 220 starless cores have been observed in the CS (2--1) or 
N$_2$H$^+$ (1--0) lines. With a detection criterion of 
$\rm T_A^* \ga 5\sigma$, 163 out of 196 sources observed in CS (2--1) 
were detected, and 72 out of 142 sources were detected in N$_2$H$^+$ (1--0).
The ranges of line 
intensities ($\rm T_A^*$) and widths (FWHMs) are approximately  
$0.05 - 0.4$ K and $\rm 0.3 - 0.9~km~s^{-1}$ for CS 
in about $80\%$ of the detected sources,  and  
$0.05-0.4$ K and $\rm 0.2-0.55~km~s^{-1}$ for the N$_2$H$^+$ isolated 
component in about $90\%$ of the detected sources.
The peak intensity and FWHM of the CS line for the detected sources 
have mean and 
standard deviation of about 0.28$\pm$0.17 K and 0.68$\pm$0.29 km s$^{-1}$,
and the corresponding values for N$_2$H$^+$ are 0.19$\pm$0.11 K 
and 0.35$\pm$0.14 km s$^{-1}$.
Table 1 lists the source coordinates and velocities for the cores  
which have been observed in CS (2--1) or N$_2$H$^+$ (1--0), 
but detected only in one line or not detected in either line. 

As a result of our survey, 69 starless cores were detected in both CS (2--1) 
and N$_2$H$^+$ (1--0), and these sources are suitable for a comparison 
between the CS and $\rm N_2H^+$ line velocities.  Table 2 lists their 
coordinates,  line velocities, $\rm N_2H^+$ line width and
optical depth, and the line asymmetry parameter  $\delta V_{CS}$ which 
will be explained in $\S 3.4$. 

The CS peak velocities and their 1$\sigma$ uncertainties in 
Tables 1 and 2 were determined from Gaussian fits.
CS spectra consisting of two components or of a brighter component 
with a shoulder were fitted only for the brighter component by masking 
the fainter part or the shoulder in order to obtain the velocity of the 
peak component of the spectrum.  The values of LSR velocity, FWHM, 
and their uncertainties for N$_2$H$^+$ (1--0) were obtained
by fitting seven Gaussian hyperfine components with the line 
parameters (velocity offsets and relative intensities) as 
determined by Caselli et al. (1995), using the Gaussian hyperfine structure 
(hfs) fit routine in the `CLASS' reduction program (Buisson et al. 1994).

In Figure 1, we show 43 pairs of CS (2--1) and N$_2$H$^+$ (1--0) 
(isolated component) spectra for sources with strong detections, 
and in Fig. 2 we present similar pairs for sources 
whose isolated components were barely detected or not seen.
In this second group, the six less-isolated hyperfine components of the 
N$_2$H$^+$ line are usually bright enough to determine the peak velocity 
of the N$_2$H$^+$ line with useful precision, and the velocity is 
indicated in Fig. 1 and Fig.  2 by a dashed vertical line.
We also present, in Fig. 3, spectra of 9 sources that show interesting 
non-Gaussian features in the CS line, 
while the N$_2$H$^+$ line was not observed or not detected. 

We observed 30 sources in C$^{18}$O (1--0) and  all were detected. 
The C$^{18}$O spectra for 5 sources which
have not been detected in the N$_2$H$^+$ (1--0) line are shown in Fig. 3.
These C$^{18}$O lines help to examine possible 
velocity shifts in the CS spectra.

\subsection{Analysis of the CS Lines}
As illustrated in Figs. 1, 2, and 3,
a substantial fraction of the CS(2--1) spectra show 
non Gaussian features,  with some type of asymmetry like
two peaks or a peak with shoulder.
This is in contrast with what is seen in 
$\rm N_2H^+$, where the isolated component appears as a
single Gaussian feature, narrower than the equivalent
CS line. This suggests that the CS asymmetries may be related 
to the higher optical depth of this line, and that they 
probably arise from self absorption by lower excitation foreground gas.
Furthermore, we notice that the spectral shapes of the CS(2--1) spectra 
vary from source to source, indicating that the characteristics of the 
absorbing gas vary strongly from one object to the other.

We classify the CS spectra in 6 different groups depending on their shape 
(Fig.4). The line characteristics and typical examples for each group 
are as follows;

1.  Two peaks with the blue peak brighter than the red (Fig. 4a)
- e.g., L1521F, L183B, L1689B, L694-2, and 
L1445\footnote[1]{It is difficult to decide from our data
whether in this source 
the double peak originates from self absorption because the
N$_2$H$^+$ line was not detected. Our recent NRAO-12m 
survey (Lee, Myers, \& Plume 1999) shows that the peak velocity of 
another optically thin tracer, DCO$^+$(2-1), lies between the two CS(2--1) 
peaks}

2.  Blue peak and red shoulder (Fig. 4b) 
- e.g., L1355, L1498, L1551S-2, TMC2, B18-3, CB23, L1622A-2, 
L1696B, L158, L234E-1, L492, L1155C-1, and L1197.

3. Single symmetric peak (Fig. 4c) - e.g., L1333, L1400A, L204C-2, 
L1517A, L1517B, L462-1, and L1063.
 
4. Red peak and blue shoulder (Fig. 4d) - e.g., L1521E

5. Two peaks with the red peak brighter than the blue peak (Fig. 4e) - e.g.,  
CB246-2

6. Two peaks  with similar brightness (Fig. 4f) 
- e.g., L1521B, L1517C-1, L1544, L204C, and L429-1 

\subsection{Analysis of the N$_2$H$^+$ Lines}

The N$_2$H$^+$ (1--0) line consists of 7 hyperfine components of 
different optical depths, being 
the isolated component ($\rm F_1F=01-12$) usually optically thin
(e.g., Caselli et al. 1995).
Fig. 5 shows four types of N$_2$H$^+$ spectra seen in our survey. 
The overwhelming majority ($\sim 53$ ) of the detected N$_2$H$^+$ lines 
have Gaussian shape, as shown in Fig. 5a for L183. A few
sources, however, have double-peaked spectra, even in the isolated component,
as in L492 (Fig. 5d).
For $\sim 15$ sources, the isolated component is too weak to be 
detected, while the
other six components of the spectrum are fairly bright (e.g., Fig. 5b). 
For these sources, the six less-isolated components play an important 
role in determining the peak velocity of N$_2$H$^+$ (1--0) line. 
The spectral shapes of several sources are sometimes different from 
component to component.
We find that 6 sources (L1521F, L1551S-2, L1544, L63, L429-1, and L492) have 
double-peaked lines in the several hyperfine components 
(e.g., Fig5-c, and d). 
Our recent surveys in $\rm H^{13}CO^+~(1-0)$ (Lee, Myers, \& Park 1999) 
and $\rm DCO^+~(2-1)$ (Lee, Myers, \& Plume 1999) 
show that all these sources present single Gaussian components 
in other thin tracers
[$\rm H^{13}CO^+~(1-0)$ and $\rm DCO^+~(2-1)$], indicating 
that the two  N$_2$H$^+$  peaks arise not from overlapping components, 
but from self-absorption due to high optical depth. 
It is noted that L1521F and L1544 
show a blue asymmetry in the main component ($\rm F_1F=23-12$) of the 
N$_2$H$^+$ line.

\subsection{Distribution of the Velocity Differences between 
CS (2--1) and N$_2$H$^+$ (1--0)}

In a previous section we have classified qualitatively the 
CS line profiles into 6 groups relying on their spectral shapes.
Now we quantify this by estimating the amount by which 
the CS (2--1) spectrum
is blue- or red-shifted with respect to the N$_2$H$^+$ (1--0) line.
To do this, we follow M97 and use the normalized velocity difference 
$ \delta V = (V_{CS}-V_{N_2H^+})/\Delta V_{N_2H^+}$, where 
$V_{CS}$ and $V_{N_2H^+}$ are the peak velocities of 
the CS and N$_2$H$^+$ lines, 
and $\Delta V_{N_2H^+}$ is the FWHM of N$_2$H$^+$ (1--0).
The calculated values of $\delta V_{CS}$ for the 
69 sources in our survey with both CS and  N$_2$H detections 
are listed in Table 2.
However, for further analysis, we drop 2 sources (L1544 and L429-1) whose
intensity ratio between the blue component ($\rm T_b$) 
and the red component ($\rm T_r$)
of the CS spectra is not significantly different from unity  
[$\rm {T_b\over T_r}\la 1+\sigma_{({T_b\over T_r})}$],
so any choice of a brighter component is ambiguous.  

In Fig. 6 we show two histograms of the distribution of $\delta V_{CS}$ in
our sample. The left panel represents the distribution for the set of 
frequencies recommended by Tafalla et al. (1998), which  
corresponds to the N$_2$H$^+$ (1--0) frequency determined by 
Caselli et al. (1995) and
the CS (2--1) frequency recommended by Pickett et al. (1998). The right panel
represents the same distribution when we use Lovas (1992) recommended 
value for CS (2--1) and keep the same value for N$_2$H$^+$. 
In the Appendix we discuss in detail
why the first frequency set should be preferred, although here we will
discuss both options in parallel. Probably the strongest conclusion
from Fig. 6 is how crucial (and urgent) are high-precision
frequency measurements for future infall searches.

Adopting the first (and preferred) set of frequencies, the $\delta V_{CS}$
distribution is clearly skewed to the blue side ($\delta V<0$), with 
a mean value ($\pm$ standard error of the mean) for the 67 sources of
$-0.24\pm 0.04$. This number, very close to 
the $-0.28\pm 0.10$ found by M97 for a set of Class 0 sources
with broader lines, 
suggests that inward motions play an important role in the kinematics
of starless cores (see below).
A statistical t-test with unknown standard deviation 
shows that the probability of drawing the observed $ \delta V_{CS}$ 
distribution from a zero mean t-distribution is less than 5$\%$, so 
the skewness is highly significant.

In Table 3, we have marked  a `B' for the 20 sources with a significant
blue asymmetry ($\delta V_{CS} \le -5~\sigma_{\delta V_{CS}}$), 
and a `R' for the 3 sources with significant red asymmetry
($\delta V_{CS} \ge 5~\sigma_{\delta V_{CS}}$).

If we adopt the second set of frequencies, however, the distribution
appears symmetric, with a mean indistinguishable from zero, and a high
probability (about 80\% in the t-test) of arising from a symmetric
population. Until higher precision CS (2--1) frequencies are available,
no further progress seems possible in determining
the statistics of  $ \delta V_{CS}$ with more accuracy.

A more robust result, as it holds for the two possible frequency choices,
is the systematic decrease of
$\delta V_{CS}$ as a function of the N$_2$H$^+$
(1--0) total optical depth $\tau_{N_2H^+}$. To study this,
we have measured
$\tau_{N_2H^+}$ using the hyperfine structure (hfs) fitting 
routine in CLASS (Buisson et al. 1994) with the relative frequencies
of the components as determined by Caselli et al. (1995), and the
derived values are indicated in Table 2. The error in
$\tau_{N_2H^+}$ can become unacceptably large for very high or very 
low optical depths, or if the N$_2$H$^+$ (1--0) spectrum is too noisy, 
and for that reason, we have discarded 13 sources for most of which 
the estimated $\tau_{N_2H^+}$ error is larger than the measurement 
(these sources have an `*' before their $\tau_{N_2H^+}$ values in Table 2).

Fig. 7 illustrates the distribution of $\delta V_{CS}$ as a function 
$\tau_{N_2H^+}$ for our preferred set of
frequencies (see Appendix), although the same trend is found if the 
Lovas (1992) value is adopted. Fig. 7 shows that there is a strong tendency
for $\delta V_{CS}$ to become more negative with increasing 
total optical depth, and all 6 sources with 
$\tau_{N_2H^+}$ larger than 11 have large negative $\delta V_{CS}$.
A similar trend is also seen in Fig. 8 which shows 
$\delta V_{CS}$ as 
a function of the total integrated intensity of the $\rm N_2H^+$ line --
a measure of the N$_2$H$^+$ column density. 
The incidence of sources with blue asymmetry ($\delta V<0$)
clearly increases with 
increasing $\rm N_2H^+$ integrated intensity.

\section{Discussion}
\subsection{Implication of the velocity difference distribution}
In the last section we have shown that if we adopt our preferred set of 
frequencies, 
the statistics of the velocity difference between $\rm CS (2-1)$ 
and $\rm N_2H^+~(1-0)$ is significantly  skewed
to the blue, and that even if we adopt the Lovas (1992) frequency for CS, 
the objects with large $\tau_{N_2H^+}$ systematically
have negative $\delta V_{CS}$. Here we
discuss the possible kinematic origin for this trends. 

Bipolar outflows can strongly affect the shape of molecular lines and
give rise to a strong asymmetric profiles even in dense gas tracers
used in our survey (e.g., M97). Our sample, however, was selected
to contain starless cores, and therefore should have no 
molecular outflows. To test this assumption, 
we have examined all spectra looking for high velocity 
wings. Among 163 CS spectra, only two sources (L183B and L429-1) show  
small wing components, in agreement with our expectation that
our sources are truly starless and have no detectable signs of
star formation. We therefore conclude that
outflow contamination is not affecting our statistics of 
the velocity difference distribution.

Core rotation can also produce blue and red spectral
asymmetry at either side of the rotational axis 
(Adelson \& Leung 1988, Zhou 1995).
Our recent mapping survey in $\rm CS(2-1)$ and $\rm N_2H^+~(1-0)$ 
(Lee, Myers, \& Tafalla 1999), however,
shows that the incidence of sources with a well-defined shift from 
blue to red asymmetry
is also small, suggesting that core rotation does not affect 
our statistics of velocity difference distribution either.

	Finally, inward motions (infall) 
will also give rise to blue asymmetry in 
spectral profiles. Line excitation in a centrally 
concentrated core will tend to
decrease outwards (unless compensated by a strong temperature gradient),
so the foreground gas will appear in absorption in optically
thick lines. If inward motions prevail, the foreground absorbing 
material will be
red shifted (approaching the background), so the self absorption will
preferentially affect the red side of the line and the resulting 
profile will have a brighter blue peak (e.g., Leung \& Brown 1977). Inward
motions (if spherical) have the property of systematically 
producing blue asymmetry in the spectra, in contrast with bipolar outflow
and rotation, which produce both blue and red asymmetries in different parts
of the core. This suggests that even if outflow and rotation were
present, when studying a large sample of objects like ours, their effect will
cancel out, and if infall is present, it will dominate the statistics
of spectral asymmetries (see also M97). We therefore conclude that the most
likely explanation for the overabundance of blue asymmetric starless
cores in our sample (if our frequency choice is correct), and the 
overabundance of blue asymmetric cores at the highest $\rm N_2H^+~(1-0)$
optical depths (independent of the frequency choice) is caused by the
presence of inward motions in the gas.
 
We emphasize that the above conclusion refers to a statistical property 
of the sample as derived from 
single-pointing observations, and it does not guarantee that each source
with blue asymmetric profile has necessary inward motions over a
significant range of gas density and spatial extent. 
Spatial mapping of the spectral asymmetry is still necessary to verify that
a given potential infall candidate is a truly infalling core. 

\subsection{Infall Candidates}
One important objective of this paper is identifying infall candidates for 
further, more detailed study. To carry out this identification, 
we analyze the following three spectral properties: 
(1) value of $\delta V_{CS}$, (2) intensity ratio of the blue 
to the red component in double-peaked
CS spectra, and (3) intensity ratio of the blue to the red 
component in double-peaked N$_2$H$^+$ spectra.
The result of this analysis is summarized in Table 3 (for our
frequency choice, see Appendix). 

Property (1) is indicated as `B', `N', or `R' 
in the second column of Table 3 depending on the value of $\delta V_{CS}$
(`B' for $\delta V_{CS} \le -5\sigma_{V_{CS}}$, `N' for 
$\vert \delta V_{CS}\vert < 5\sigma_{V_{CS}}$,
and `R' for $\delta V_{CS} \ge 5\sigma_{V_{CS}}$). 
To be a good infall candidate, a source should have a `B' in this column. 

Property (2) can be also expressed as `B', `N', or `R' according to the ratio
$\rm T_b/T_r$ (`B' for $\rm T_b/T_r > 1+\sigma$, 
`N' for $\rm 1-\sigma \le T_b/T_r \le
1+\sigma$, and `R' for $\rm T_b/T_r < 1-\sigma$), where 
the values of $\rm T_b$ and $\rm T_r$ were derived from
Gaussian fits to
each component of the CS spectra with double peaks or single
peak  with distinct shoulder. To fulfill our criterion of being a 
good infall candidate, the ratio  should be greater than 1.0 within 
a 1$\sigma$ error.
This criterion allows us include sources whose $\delta V_{CS}$s  are 
poorly determined due to weak $\rm N_2H^+$ 
emission, but  whose CS spectra show an 
apparently self-absorbed infall asymmetry (e.g., L1689B).
 
Property (3) can also help identifying infall candidates,
because the  N$_2$H$^+$ hyperfine components  trace higher density 
gas (a few $10^5$ cm$^{-3}$) than CS (2--1) (a few $10^4$ cm$^{-3}$), 
when their different optical depths are taken into account. 
Like with property (2), we have marked `B',  `N', or `R' those
sources satisfying 
$\rm T_b/T_r > 1+\sigma$, $\rm 1-\sigma \le T_b/T_r \le 1+\sigma$, or 
$\rm T_b/T_r < 1-\sigma$, respectively.

The above properties used for selecting infall candidates are similar 
to those used by M97 and Gregersen et al. (1997). Given their multiplicity,
we can define different quality degrees for the candidates.  Thus, we
define as a ``strong'' infall candidate a source that is blue (B) 
according to at least 2 criteria, and is 
not red (R) according to any. A 
``probable'' infall candidate is defined as one 
which is blue (B) according to one criterion and is not
red (R) according to any.
With these rules, we find 7 ``strong'' infall 
candidates --{\bf L1355, L1498, 
L1521F, L183B, L158, L694-2, and  L1155C-1}, and 10 
``probable'' infall candidates --
{\bf L1524-4, L1445, TMC2, B18-3, CB23, 
L1544, L1622A-2, L1689B, L234E-1, and L492}.
	
For the list of ``probable'' infall candidates, we have dropped  several 
sources that are `B' according to property (1)  
due to a very narrow $\rm N_2H^+$ line, but that do not show 
any pronounced asymmetry in the spectral profiles. These sources
are L1400K, TMC1, L183, L1696A, L922-2, L1155C-2, and L1234. Two
sources in our infall-candidate list, 
L1544 and L1498, are previously known and well studied infall 
candidates (see Tafalla et al. 1998,
Williams et al. 1999 for L1544; Kuiper, Langer, \& Velusamy 1996, 
Willacy, Langer, \& Velusamy 1998 for L1498), but the additional
15 sources in our list are new infall candidates.
In contrast, we have found 1 strong outflow candidate (CB246-2)
and one probable outflow candidates (L63), according to the 
equivalent rules as those described before
(L1049-1 is marked with ``R'' according to the spectral property (1),
but it does not show any pronounced asymmetry in the CS spectrum, so 
we have excluded it from the
category of ``probable'' outflow candidate). 

All the above analysis has been done using our preferred set of frequencies
(see Appendix). If the Lovas (1992) CS (2--1) frequency is used instead,
the number of infall candidates decreases, but  
all 7 `strong' infall candidates still remain `strong'.
Five of the 10 `probable' infall candidates, however,
become `N,' while 5 other sources (L1445, TMC2, L1544, L1689B, \& 
L492) remain `probable'.

\subsection{Infall Kinematics of The Starless Cores}
Typical infall speeds for our candidates 
can be approximately calculated  by taking half of the velocity dispersion
($={1\over2}\sigma_v= {1\over 2}~{\Delta V_{N_2H+} \over (8~ln2)^{1\over2}}$)
of the optically thin $\rm N_2H^+$ line (e.g., Leung \& Brown 1977, 
Myers et al. 1996).
The infall speeds of all sources are found to be approximately 
between $\rm 0.04 - 0.1~km~s^{-1}$. 

For each core, we have calculated the line FWHM for thermal motions 
[$\Delta V_T=(8~ln2~kT/m)^{1/2}$,
where $k$ is the Boltzmann constant, $T$ the kinetic temperature, and 
$m$ the mean molecular
weight (2.3 amu)], and the line FWHM for non-thermal motions 
[$\Delta V_{NT}=(\Delta V_{N_2H^+}^2 - {8~ln2~kT\over m_{N_2H^+}})^{1\over2}$, 
where $\rm m_{N_2H^+}$ is the molecular
weight of $\rm N_2H^+$ (29 amu)] using the observed $\rm \Delta V_{N_2H^+}$
and assuming $\rm T=10~K$. Taking the ratio of the two
($\Delta V_{NT}/ \Delta V_T$), we find that it is always smaller than 1 
except for one source (L492; ratio equal to 1.13), meaning that
most of the starless infall candidates are in state of 
`thermal' infall. This
is contrast to the result of M97, who found a prevalence of turbulent
infall motions when studying a sample of cores with embedded YSOs. We
note that this difference arises from the broader lines in the M97
sample, and not from their  $\delta V_{CS}$ value, which is very
similar to that found here for starless cores.
 
\subsection{Future Work}

Though our statistical analysis of the distribution of 
velocity differences strongly suggests that inward motions 
dominate the kinematics in starless cores, it is still premature to state
that the blue asymmetry in each of our candidates is purely from
inward motions, and it is not contaminated by other elements 
like outflow or rotation.  In addition, even if the motions we observe
are truly inward, it can not be decided yet whether they arise from
`inside-out' gravitational collapse 
(Shu 1977), or from contraction due to ambipolar diffusion 
(Mestel \& Spitzer 1956) or any other process [e.g., 
pressure-driven motions (Myers \& Lazarian 1998)].

A more comprehensive picture of the kinematics of starless cores 
requires further observations of additional transitions
[e.g., $\rm CS~(3-2)$], different gas tracers (e.g., ions such as 
$\rm HCO^+$), and 2-dimensional mapping in both lines and continuum.
Work in the above directions is currently in progress.

\section{Conclusions}

We have carried out a survey of 220 starless cores by observing them
with the  Haystack 37-m telescope
in the optically thick CS (2--1) line, and the optically thin
N$_2$H$^+$ (1--0) and C$^{18}$O (1--0) lines.
The main goal of this survey has
been to study the presence and frequency of inward motions in 
cores which have not yet formed stars. 

The sources for these observations were selected primarily from 
the catalogue of optically selected cores by LM99, although a 
few new sources were added and several positions were slightly changed.
Out of 227 sources in the LM99 catalogue observable from the latitude of 
the Haystack Observatory, about 93$\%$ (212) have been observed.

As a result of the survey, 163 out of 196 sources observed were detected in CS, 
72 out of 142 were detected in N$_2$H$^+$, and  30 out of 30 
were detected in C$^{18}$O.
In total, 69 sources were detected in both CS and N$_2$H$^+$, and these
are the sources used for our analysis of inward motions.

The main conclusions of this analysis are:

1. The CS lines change in shape strongly from one source to the other, and
very frequently present asymmetric features.
From a comparison with the optically thin lines of $\rm N_2H^+$ (1--0) 
and C$^{18}$O (1--0), which are usually symmetric and Gaussian, we conclude
that most of the CS asymmetries arise from a combination of optical
depth effects and kinematics. We classify the CS spectra
according to their shape in the 6 following groups:
(1) two peaks, with  blue peak brighter than red,
(2) blue peak and red shoulder, (3) single symmetric peak, 
(4) red peak and blue shoulder, (5) two peaks, with red peak brighter 
than blue peak, 
and (6) two peaks with similar brightness.
 
2. The isolated component of N$_2$H$^+$ (1--0) ($\rm F_1F=01-12$) 
is found to be optically thin in most sources.
For 6 sources, however,
several hyperfine components of the N$_2$H$^+$ line show 
what appear to be self absorption features 
that differ from component to component.
L1521F and L1544 show blue asymmetry in
the main component of the N$_2$H$^+$ line ($\rm F_1F=23-12$).

3. We find consistent indications that inward motions are 
a significant feature in the starless cores of our large sample.

First, we find that the distribution of the normalized velocity difference 
$ \delta V_{CS}$ between the CS and N$_2$H$^+$ lines 
for our sample is significantly skewed to the blue side ($\delta V_{CS} < 0$). 
The statistics of the average value of $\delta V_{CS}$ and 
of the student's t-test are similar to 
those found for cores with 
embedded Class 0 sources (Mardones et al. 1997).
There is also an excess of sources (20) with 
$\delta V_{CS} \le -5~\sigma_{\delta V_{CS}}$  over
(3) sources with $\delta V_{CS} \ge 
5~\sigma_{\delta V_{CS}}$. In addition,
the incidence of the sources 
with infall asymmetry tends to increase
as the total optical depth or the integrated intensity  
of the N$_2$H$^+$ line increases.

Second, among sources with self-absorbed features in the CS line,
there is an excess of sources (10) for which the 
intensity ratio between the blue and the red peak is larger than 1.0,
while there is just one source where the ratio is less than 1.0.

5. Based on three spectral properties -- the value of $\delta V_{CS}$, 
the intensity ratio of the blue to the red peak  in two-peaked
CS spectra, and the intensity ratio of the blue to the red peak 
in two-peaked main hyperfine components of N$_2$H$^+$ (1--0),
we identify  7 {\it strong infall candidates} --{\bf L1355, L1498,
L1521F, L183B, L158, L694-2, and  L1155C-1},
and 10 {\it probable infall candidates}--
{\bf L1524-4, L1445, TMC2, B18-3, CB23, L1544, L1622A-2, 
L1689B, L234E-1, and  L492}.
All sources but two (L1498 and L1544) are newly found from this survey. 

6. If the motions we observe are true infall, their typical speed is 
about $\rm 0.04 - 0.1~km~s^{-1}$. This implies infall in starless
cores is subsonic and so in thermal motion, in contrast with the turbulent infall
speeds found by M97 in cores with embedded YSOs.

7. Our identification of strong infall candidates and the increase
in infall asymmetry with line optical depth are independent 
of the choice of CS and $\rm N_2H^+$ line frequencies. 
The statistical distribution
of infall asymmetry, however, is crucially dependent on these frequencies. 
Higher precision, laboratory measurements of
the line frequencies of these and other dense gas tracers
are urgently needed for future progress in infall research.

\acknowledgments
This survey would have been impossible without the dedicated support 
of the staff of Haystack Observatory, among whom we would like to give 
special thanks to John Ball, Phil Shute, 
Joseph Carter, Kevin Dudevoir, and Mike Titus. 
Especially we would like to thank Phil Shute for his considerable help 
through our 2 years of observations.
We are also grateful to the Director, Joe Salah for his outstanding 
support of our program. We wish to thank Tyler Bourke for his help 
in the beginning of the survey and Diego Mardones
for many useful discussions.  
C.W.L acknowledges the financial support by 1996 Overseas 
Postdoctoral Support Program  of 
Korea Research Foundation and partial support by 
Star Project Grant 98-2-400-01 of the 
Korea Astronomy Observatory. C.W.L also thanks 
the Harvard-Smithsonian Center for Astrophysics
for support while working on this project. 
M.T acknowledges partial support from Spanish DGES grant PB96-104.
This research 
was supported  by NASA Origins of 
Solar System Program, Grant NAGW-3401.

\appendix
\label{boloappendix}
\section{Selection of the Line Frequencies}
Tafalla et al. (1998) discussed how the narrow lines observed towards the
starless core L1544 could be used to check the consistency of the different
sets of recommended line frequencies, and they proposed a 
set of values that seemed self-consistent when checked against 
observations of that core.
The results of the L1544 analysis from Tafalla et al. (1998) were
independent from the exact frequency values, so the frequency discussion 
was stated 
without further illustration. The work presented in the
present paper, however, depends sensitively on the exact frequencies
adopted for the lines (see Fig. 6), so it seems necessary to provide
further details on how the frequency comparison was done. 

To minimize systematic errors, it is convenient to use a set of observations
done with the same telescope using a similar spatial resolution and 
sampling, so
for our discussion we will use only FCRAO QUARRY data (a similar analysis
using IRAM 30m data leads to the same conclusions). To increase the S/N
of the spectra, necessary for the weak lines, 
we average the spectra over a FCRAO QUARRY 
footprint ($\sim 5'\times 4.2'$ area), and show the results in Fig. 9.
The first panel of this figure shows the result of using the frequencies
recommended by Lovas (1992) [109782.160 MHz for $\rm C^{18}O~(1-0)$, 
96412.982 MHz for $\rm C^{34}S~(2-1)$, and 97980.968 MHz for
$\rm CS~(2-1)$]. As
Fig. 9-(a) clearly shows, unless the velocity shifts between the 
different tracers are real, which is very unlikely given their
similar line widths and spatial distributions, the frequencies 
are inconsistent. 
 
Next (Fig. 9-(b)), we use the set of frequencies 
recommended by Lovas \& Krupenie (1974), 
one of the original references for the CS and CO frequencies 
in Lovas (1992) catalogue. These values are 109782.182 MHz for 
$\rm C^{18}O~(1-0)$,
96412.953 MHz for $\rm C^{34}S~(2-1)$, and 97981.007
MHz for $\rm CS~(2-1)$ and produce
an excellent agreement between the $\rm C^{18}O$ 
and $\rm C^{34}S$ spectra. 
However, there is still a clear discrepancy with $\rm CS~(2-1)$, which seems 
slightly displaced to the red.
 
To correct the CS discrepancy, Tafalla et al. (1998) chose 97980.950 MHz for 
the $\rm CS~(2-1)$ frequency, which is the value that Pickett et al.
(1998) recommend, and as shown in Fig. 9-(c), produces a good
agreement with the other lines.

As a further check, Fig. 9-(d) presents a comparison of the
$\rm C^{18}O~(1-0)$ and $\rm C^{34}S~(2-1)$ lines with the isolated
$\rm N_2H^+~(1-0)$ component ($\rm F_1F=01-12$), whose frequency was
determined with high accuracy by Caselli et al. (1995) (93176.2651 MHz,
a value we have adopted for this work). As the figure shows, 
there is an excellent agreement between the three lines, reinforcing
our belief that the frequency selection is correct.

One possible point of concern is the fact that L1544 shows
infall asymmetry, and that this may in some way bias the 
adopted frequency to be higher than the real value. 
If this was the case, however, one would
expect that the bias would produce in non infalling objects
an opposite signature, and would give rise to a false overabundance of 
outward motions in a statistical study like ours. As shown
in Fig. 6, the opposite is the case, making us to believe  that
the infall nature of L1544 is immaterial for the frequency determination.

Finally, we examine the consistency of our frequencies 
by comparing the  C$^{18}$O (1--0) peak velocities  
with  those of N$_2$H$^+$ (1--0) for 16 sources 
whose C$^{18}$O (1--0) spectra are Gaussian and have good S/N.
The difference ($\rm V_{N_2H^+} - V_{C^{18}O}$) is negligible 
(about  $-0.02\pm 0.057$ km s$^{-1}$), again in agreement with 
our expectations.

In summary, the frequencies recommended here 
[97980.950 MHz for the $\rm CS~(2-1)$, 
93176.2651 MHz for the $\rm N_2H^+~(1-0)$ isolated component, and 
109782.182 MHz for the $\rm C^{18}O~(1-0)$] give the best agreement 
among the lines observed toward L1544.  We note, however, that 
our comparison is a relative one, so it is possible the the
absolute frequency values may be slightly off. If this is
the case, all frequencies would have to be changed accordingly,
and the results in this paper would not be affected. 

\vfill\eject

\clearpage

\clearpage
\begin{figure}
\begin{center}
{\bf FIGURE CAPTIONS}
\end{center}
\end{figure}

\begin{figure}
\noindent{\bf Fig. 1. ---} CS (2--1) and N$_2$H$^+$ (1--0) 
(isolated component) line spectra for 43 sources
with strong detections. The dashed line indicates 
the velocity of N$_2$H$^+$ obtained
from the Gaussian hfs fit of its 7 hyperfine components.
\end{figure}

\begin{figure}
\noindent{\bf Fig. 2. --- } 
CS (2--1) and N$_2$H$^+$ (1--0) (isolated component) spectra for 21 sources
with weak or no detection of the N$_2$H$^+$ isolated component, but
for which the other N$_2$H$^+$ components
are bright  enough to determine a velocity from the hyperfine structure fit.
\end{figure}

\begin{figure}
\noindent{\bf Fig. 3. --- } Spectra for 9 selected sources
with non Gaussian CS lines for which 
N$_2$H$^+$ was not detected or not observed. For 5 sources,
C$^{18}$O (1--0) spectra are also presented, and their central 
velocities are indicated by dashed lines.
\end{figure}

\begin{figure}
\noindent{\bf Fig. 4. --- } Different types of CS spectra -
(a) Two peaks, with  blue peak brighter than red,
(b) Blue peak and red shoulder,
(c) Single symmetric peak,
(d) Red peak and blue shoulder,
(e) Two peaks, with red peak brighter than blue peak, and
(f) Two peaks  with similar brightness.
Dashed lines are the Gaussian fit velocities of the optically 
thin $\rm N_2H^+~(1-0)$ or $\rm C^{18}O~(1-0)$.
\end{figure}

\begin{figure}
\noindent{\bf Fig. 5. --- } Representative spectra 
of N$_2$H$^+$ (1--0) - (a) 
bright line with no self absorption features 
in any of the hyperfine components,
(b) isolated component barely detectable while
the  other hyperfine components
are bright enough to determine the LSR 
velocity of the N$_2$H$^+$ line,
(c) symmetric isolated component but possible self absorption in the
other hyperfine components,
(d) self-absorption in the 
isolated component as well as in the other components.
The dotted lines in each panel mark the velocities of the
7 components of the N$_2$H$^+$ obtained
from a Gaussian hyperfine structure (hfs) fit using CLASS 
(Buisson et al. 1994) with 
the velocity offsets determined by Caselli et al. (1995).
A Gaussian hfs fit spectrum is overlaid on 
the observed spectrum of L204C-2 
in panel (b). 
\end{figure}

\begin{figure}
\noindent{\bf Fig. 6. ---} Histograms of the normalized velocity difference
($\delta V_{CS}$) between V$_{CS(2-1)}$ and V$_{N_2H^+(1-0)}$ for a sample
of 67 objects with detection in both tracers. The left panel represents the
distribution using our preferred set of frequencies (see Appendix for 
a full discussion on frequencies), and shows 
a significant excess of blue shifted sources ($\delta V<0$). The right
panel shows the histogram we would get if the CS (2--1) frequency from 
Lovas (1992) is used and the $\rm N_2H^+$ (1--0) 
frequency is not changed. 
\end{figure}

\begin{figure}
\noindent{\bf Fig. 7. ---} 
Distribution of $\delta V_{CS}$ as a function of total optical depth 
($\tau_{N_2H^+}$) of the N$_2$H$^+$ (1--0) line.
The $1 \sigma$ error bars drawn in the Figure are the means of 
the errors in $\tau_{N_2H^+}$ and
$\delta V_{CS}$, respectively.
\end{figure}

\begin{figure}
\noindent{\bf Fig. 8. ---} 
Distribution of $\delta V_{CS}$ as a function of 
total integrated intensity of the $\rm N_2H^+$
hyperfine components.
\end{figure}

\begin{figure}
\noindent{\bf Fig. 9. ---} 
Comparison of $\rm C^{18}O~(1-0)$
(thick solid spectra), $\rm CS~(2-1)$ (thin solid spectra), 
$\rm C^{34}S~(2-1)$
(dotted spectra ), and $\rm N_2H^+~(1-0)$ (dashed spectrum) towards L1544
(average over a FCRAO QUARRY footprint, see text).
(a) Lovas (1992) frequencies for CS (2--1), $\rm C^{34}S~(2-1)$,
and $\rm C^{18}O~(1-0)$,
(b) Lovas \& Krupenie (1974) frequencies
for CS (2--1), $\rm C^{34}S~(2-1)$, and $\rm C^{18}O~(1-0)$,
(c) Pickett et al. (1998) frequency  $\rm CS~(2-1)$, and
Lovas \& Krupenie (1974) frequencies for $\rm C^{34}S~(2-1)$ and
$\rm C^{18}O~(1-0)$, and
(d) Caselli et al. (1995) frequency for $\rm N_2H^+~(1-0)$, 
Lovas \& Krupenie (1974) frequencies for $\rm C^{18}O~(1-0)$ and
$\rm C^{34}S~(2-1)$. The intensities of all the 
spectral lines are scaled up to be the same
as the intensity of $\rm C^{18}O$. 
\end{figure}


\begin{references}

\reference{ } Adelson, L.M. \& Leung, C.M. 1988, MNRAS, 235, 349

\reference{ } Ball, J.A., Carter, J.C., Shute, P.A., 
Philips, A.E.E., Whalen, R.D. 
\& Johnson, H.N. 1993, in An Introduction to the Haystack Observatory

\reference{ } Barvainis, R., Ball, J.A., Ingalls, R.P. \& Salah, E.J. 1993, 
PASP, {105}, 1334

\reference{ } Beichman, C. A., Myers P.C., Emerson, J. P., Harris, S., 
Mathieu, R., Benson, P.J., \& Jennings, R . E., 1986 \apj,\ {307}, 337

\reference{ } Benson P.J., \& Myers P.C. 1989, \apjs, {71}, 89

\reference{ } Buisson, G., Desbats, L., Duvert, G., Forveille, T., 
Gras, R., Guilloteau, S., 
Lucas, R., \& Valiron, P. 1994, CLASS Manual (Grenoble: IRAM)

\reference{ } Ciolek, G.E., \& Mouschovias, T.Ch. 1995, \apj, {454}, 194

\reference{ } Caselli, P., Myers, P.C., \& Thaddeus, P. 1995, \apj, {455}, L77

\reference{ } Gregersen, E.M., Evans, N. J., Zhou, S., \& Choi, M. 1997, 
\apj,  {484}, 256

\reference{ } Kuiper, B.H., Langer, W.D., \& Velusamy, T. 1996, \apj, 468, 761

\reference{ } Lee, C. W., \& Myers, P.C. 1999, \apjs, in press (LM99)

\reference{ } Lee, C. W., Myers, P.C., \& Park, Y.S. 1999, in preparation

\reference{ } Lee, C. W., Myers, P.C., \& Plume, L. 1999, in preparation

\reference{ } Lee, C. W., Myers, P.C., \& Tafalla, M. 1999, in preparation

\reference{ } Leung, C.M., \& Brown, R.B. 1977, ApJ, 214, L73

\reference{ } Lovas, F.J. 1992, J. Phys. Chem. Ref. Data, {21}, 181

\reference{ } Lovas, F.J. \& Krupenie, P.H. 1974, J. Phys. Chem. 
Ref. Data, {3}, 25

\reference{ } Mardones, D., Myers, P.C., Tafalla, M., Wilner, D.J., 
Bachiller, R., \& Garay, G. 1997, ApJ, {489}, 719 (M97)

\reference{ } Mestel, L., \& Spitzer, L. 1956, MNRAS, 116, 505

\reference{ } Myers, P.C., Mardones, D., Tafalla, M., Williams, J.P., 
\& Wilner, D.J. \apjl, 1996, {465}, 133 

\reference{ } Myers, P.C., \& Lazarian, A. 1998 \apj, {507}, L157 

\reference{ } Pickett, H. M., Poynter, R. L., Cohen, E.A., Delitsky, M.L., 
Pearson, J.C., \& 
Muller, H.S.P. 1998, J. Quant. Spectrosc. \& Rad. Transfer 60, 883


\reference{ }  Shu, F.H. 1977, \apj, {214}, 488

\reference{ }  Shu, F.H., Adams, F.C., \& Lizano, S. 1987, ARAA, {25}, 23

\reference{ }  Tafalla, M., Mardones, D., Myers, P.C., Caselli, P., 
Bachiller, R., \& Benson, P.J. 1998, ApJ, 504, 900

\reference{ }  Ward-Thompson, D., Scott, P.F., Hills, R.E., 
\& Andr\'e P. 1994, MNRAS, {268}, 276

\reference{ }  Ward-Thompson, D., Motte, F., \& Andr\'e, P. 1999, 
MNRAS, in press.

\reference{ }  Willacy, K., Langer, W.D., \& Velusamy, T. 1998, 
ApJ, {507}, L171

\reference{ }  Williams, J.P., Myers, P.C., Wilner, D.J., \& Di Francesco, J. 
1999, ApJ, 513, L61 

\reference{ }  Zhou, S., Evans, N.J., Kompe, C., \& Walmsley, C. M. 
1993, ApJ, 404, 232

\reference{ }  Zhou, S. 1995, ApJ, 442, 685.

\end{references}
\end{document}


\addtocounter{page}{20}
\pagestyle{headings}

\begin{deluxetable}{lrrrcc}
\small
\tablecaption{ Starless Cores observed in CS (2-1), N$_2$H$^+$ (1-0), and C$^{18}$O (1-0)
\tablenotemark{a}}
\tablehead{
        \colhead{Source\tablenotemark{b}}      &\colhead{RA 1950} & \colhead{DEC 1950}     &
        \colhead{$V_{CS}$}&\colhead{$V_{N_2H^+}$}   &\colhead{$V_{C^{18}O}$}   \nl
        \colhead{} & \colhead{$^h~~^m~~^s~$}&\colhead{~~~~$^{\circ}~~~'~~~''$} &
        \colhead{km~s$^{-1}$}&\colhead{km~s$^{-1}$}&\colhead{km~s$^{-1}$} 
        }
\startdata
  L1275     &00~00~48.2 & 67~02~05&   -6.73$\pm$0.04&         ND       & \nodata          \nl
  L1291     &00~39~47.6 & 61~44~35&    5.56$\pm$0.02&         ND       & \nodata          \nl
  L1358-2   &02~51~55.1 & 69~12~03&   -3.24$\pm$0.03&         ND       & \nodata          \nl
  B5-2      &03~45~15.5 & 32~45~07&    9.81$\pm$0.04&           \nodata& \nodata          \nl
  L1495-2   &04~14~51.2 & 28~30~40&    8.06$\pm$0.03&           \nodata& \nodata          \nl
  L1506B    &04~16~03.9 & 25~13~01&    7.53$\pm$0.01&         ND       & \nodata          \nl
  L1521B*   &04~21~08.5 & 26~30~00&    6.57$\pm$0.02&         ND       & 6.48 $\pm$0.01   \nl
  B213-7    &04~22~12.8 & 26~26~10&    6.42$\pm$0.03&         ND       & \nodata          \nl
  L1400A-2  &04~25~05.8 & 54~39~40&    3.52$\pm$0.05&         ND       & \nodata          \nl
  L1400F    &04~25~37.0 & 54~07~23&    1.66$\pm$0.08&         ND       & 1.87 $\pm$0.11   \nl
  L1521E    &04~26~09.6 & 26~07~46&    6.86$\pm$0.01&         ND       & 6.86 $\pm$0.01   \nl
  L1445     &04~28~28.8 & 46~31~01&    5.12$\pm$0.01&         ND       & \nodata          \nl
  L1551S-1  &04~28~53.9 & 18~01~14&    6.71$\pm$0.02&         ND       & \nodata          \nl
  B18-2     &04~29~34.4 & 24~45~46&    6.17$\pm$0.02&           \nodata& \nodata          \nl
  L1536-2   &04~29~50.7 & 22~53~29&    ND           &         \nodata  & \nodata          \nl
  TMC2-3    &04~29~57.6 & 24~11~26&    6.52$\pm$0.02&         ND       & \nodata          \nl
  L1536-3   &04~30~01.7 & 22~47~41&    5.69$\pm$0.10&         \nodata  & \nodata          \nl
  L1536-1   &04~30~19.7 & 22~36~51&    5.60$\pm$0.03&         \nodata  & \nodata          \nl
  L1527A-2  &04~35~54.0 & 26~15~38&    5.76$\pm$0.06&         \nodata  & \nodata          \nl
  L1507-2   &04~39~53.4 & 29~38~25&    6.26$\pm$0.01&         ND       & \nodata          \nl
  L1426     &04~43~17.5 & 52~59~03&    4.09$\pm$0.04&         ND       & \nodata          \nl
  L1517C-1**&04~51~35.9 & 30~30~50&    5.45$\pm$0.01&         ND       & 5.55 $\pm$0.01   \nl
  L1517B-2  &04~51~56.4 & 30~38~03&    5.82$\pm$0.02&         ND       & \nodata          \nl
  L1544-3   &05~00~38.6 & 25~14~18&    7.12$\pm$0.05&          \nodata & \nodata          \nl
  L1544-2   &05~01~39.1 & 25~07~30&    6.87$\pm$0.01&         ND       & \nodata          \nl
  L1523     &05~03~02.7 & 31~37~30&    7.03$\pm$0.02&          \nodata & \nodata          \nl
  CB28      &05~03~51.2 &-04~00~20&    8.74$\pm$0.01&         ND       & \nodata          \nl
  L1621-2   &05~52~49.8 & 02~08~07&    1.25$\pm$0.02&         ND       & \nodata          \nl
  L1621-1   &05~53~23.0 & 02~17~39&    1.82$\pm$0.01&          \nodata & \nodata          \nl
  CB37      &05~57~29.9 & 31~39~25&    0.92$\pm$0.09&          \nodata & \nodata          \nl
  L1574     &06~05~08.9 & 18~28~41&    0.22$\pm$0.03&          \nodata & \nodata          \nl
  L1633     &06~22~04.0 & 03~23~53&    8.93$\pm$0.03&         ND       & \nodata          \nl
  L1696B-2  &16~26~09.4 &-24~06~43&    3.18$\pm$0.02&          ND      & \nodata          \nl
  L1704-2   &16~28~09.3 &-23~34~29&      \nodata    &          ND      & \nodata          \nl
  L1757-1   &16~29~01.9 &-19~35~24&    1.45$\pm$0.07&          \nodata & \nodata          \nl
  L1757-2   &16~29~08.5 &-19~42~05&      \nodata    &          ND      & \nodata          \nl
  L1689A    &16~29~10.4 &-24~56~03&      \nodata    &          ND      & \nodata          \nl
  L1709C-2  &16~30~18.0 &-23~40~01&      \nodata    &          ND      & \nodata          \nl
  L1709C    &16~30~52.6 &-23~33~54&    2.50$\pm$0.02&          ND      & 2.57 $\pm$0.02   \nl
  L121      &16~36~41.0 &-13~59~21&     ND          &          \nodata & \nodata          \nl
  L1782-1   &16~39~13.2 &-19~36~22&    5.34$\pm$0.01&          \nodata & \nodata          \nl
  L158-2    &16~44~02.6 &-13~53~26&    3.87$\pm$0.05&          \nodata & \nodata          \nl
  L204C-1   &16~44~30.6 &-12~14~14&    4.20$\pm$0.02&          ND      & \nodata          \nl
  L204C**   &16~44~44.2 &-12~04~27&    4.89$\pm$0.02&          ND      & 4.13 $\pm$0.01   \nl
  L191-1    &16~44~59.6 &-12~34~43&    3.91$\pm$0.02&          ND      & \nodata          \nl
  L204B*    &16~45~00.0 &-11~50~00&    3.74$\pm$0.02&          ND      & 3.70 $\pm$0.01   \nl
  L204F**   &16~45~01.3 &-11~50~50&    3.79$\pm$0.02&          ND      & 3.69 $\pm$0.01   \nl
  L191-2    &16~45~01.7 &-12~47~38&    3.32$\pm$0.03&          \nodata & \nodata          \nl
  L255-2    &16~45~30.9 &-09~58~47&    3.51$\pm$0.04&          \nodata & \nodata          \nl
  L234E3    &16~45~49.0 &-10~45~45&    3.02$\pm$0.02&          ND      & \nodata          \nl
  L63-4     &16~46~49.1 &-17~49~47&     ND          &          \nodata & \nodata          \nl
  L63-3     &16~47~10.5 &-17~53~51&    5.69$\pm$0.10&          \nodata & \nodata          \nl
  L63-2     &16~47~11.2 &-17~59~52&    5.86$\pm$0.03&          ND      & \nodata          \nl
  L141-2    &16~48~03.1 &-15~16~54&    3.62$\pm$0.02&          ND      & \nodata          \nl
  L141-3    &16~48~11.1 &-15~12~04&    3.33$\pm$0.01&          ND      & \nodata          \nl
  L129-2    &16~51~57.6 &-16~21~42&    5.79$\pm$0.02&          ND      & \nodata          \nl
  L129-1    &16~52~18.4 &-16~17~10&    5.80$\pm$0.01&          ND      & \nodata          \nl
  L122      &16~52~22.4 &-16~42~03&    6.18$\pm$0.01&          ND      & \nodata          \nl
  CB68-2    &16~54~00.0 &-15~57~11&    4.99$\pm$0.02&          \nodata & \nodata          \nl
  L111      &17~11~59.5 &-20~27~15&      \nodata    &          ND      & \nodata          \nl
  L173      &17~14~25.4 &-18~27~45&      \nodata    &          ND      & \nodata          \nl
  L392      &17~50~24.3 &-08~26~43&      ND         &          \nodata & \nodata          \nl
  L421      &17~51~38.9 &-05~46~02&    7.43$\pm$0.06&          \nodata & \nodata          \nl
  L460      &17~55~24.4 &-03~46~24&    6.82$\pm$0.02&          \nodata & \nodata          \nl
  L432      &17~55~46.2 &-05~43~45&    7.37$\pm$0.03&          ND      & \nodata          \nl
  L468      &17~57~38.0 &-03~30~03&    ND           &          \nodata & \nodata          \nl
  L422      &18~09~20.0 &-08~04~46&    3.65$\pm$0.04&          \nodata & \nodata          \nl
  L438      &18~11~23.8 &-07~10~01&    4.61$\pm$0.03&          \nodata & \nodata          \nl
  L436      &18~11~53.0 &-07~23~49&    4.68$\pm$0.03&          \nodata & \nodata          \nl
  L490      &18~12~42.5 &-03~47~11&    8.09$\pm$0.04&          \nodata & \nodata          \nl
  CB130-2   &18~13~37.5 &-02~24~28&    7.17$\pm$0.04&          ND      & \nodata          \nl
  CB130-1   &18~13~39.2 &-02~34~06&      \nodata    &  7.65 $\pm$0.028 & \nodata          \nl
  CB130-3   &18~13~41.5 &-02~17~47&      \nodata    &  7.20 $\pm$0.005 & \nodata          \nl
  L328      &18~14~05.3 &-18~03~10&      \nodata    &  6.62 $\pm$0.006 & \nodata          \nl
  L429-2    &18~14~32.8 &-08~19~02&      \nodata    &          ND      & \nodata          \nl
  L429-3    &18~14~50.8 &-08~15~17&    ND           &          \nodata & \nodata          \nl
  L429-4    &18~15~05.2 &-08~10~07&    ND           &          \nodata & \nodata          \nl
  L483-2    &18~15~26.5 &-04~34~24&      \nodata    &          ND      & \nodata          \nl
  L466      &18~17~04.5 &-06~07~36&      \nodata    &          ND      & \nodata          \nl
  L539-1    &18~21~05.7 &-01~04~28&    ND           &          \nodata & \nodata          \nl
  L539-2    &18~21~46.5 &-01~01~05&    8.71$\pm$0.04&          \nodata & \nodata          \nl
  L588-2    &18~32~47.0 &-00~43~53&   10.79$\pm$0.03&          \nodata & \nodata          \nl
  L588-3    &18~33~18.2 &-00~33~25&      \nodata    &          ND      & \nodata          \nl
  L495      &18~36~32.6 &-06~41~06&      \nodata    &          ND      & \nodata          \nl
  L648-2    &18~37~20.5 & 12~47~05&    5.22$\pm$0.02&          ND      & \nodata          \nl
  L530D-3   &18~46~48.0 &-05~03~12&    3.32$\pm$0.05&          \nodata & \nodata          \nl
  L530D-1   &18~47~20.5 &-04~51~55&    3.48$\pm$0.07&          \nodata & \nodata          \nl
  L530D-2   &18~47~22.1 &-04~43~25&    ND           &          \nodata & \nodata          \nl
  L549      &18~59~29.8 &-05~23~08&      \nodata    &          ND      & \nodata          \nl
  L709      &19~11~37.3 & 16~22~01&    5.91$\pm$0.02&          \nodata & \nodata          \nl
  L763      &19~16~45.9 & 23~21~45&   11.18$\pm$0.06&          \nodata & \nodata          \nl
  L730-1    &19~17~40.5 & 19~51~09&    8.24$\pm$0.03&          ND      & \nodata          \nl
  L655      &19~17~47.2 & 08~23~33&    7.97$\pm$0.02&          \nodata & \nodata          \nl
  L730-2    &19~18~12.4 & 19~56~53&    8.10$\pm$0.05&          \nodata & \nodata          \nl
  L673-4    &19~18~47.1 & 11~20~11&    6.67$\pm$0.02&          ND      & \nodata          \nl
  L673-5    &19~18~53.6 & 11~27~16&    6.77$\pm$0.04&          \nodata & \nodata          \nl
  L673-6    &19~19~06.2 & 11~30~39&    6.67$\pm$0.04&          \nodata & \nodata          \nl
  L673-8    &19~19~15.4 & 11~12~20&    7.09$\pm$0.04&          \nodata & \nodata          \nl
  L673-10   &19~19~32.5 & 11~14~21&      \nodata    &          ND      & \nodata          \nl
  L677-1    &19~19~50.0 & 11~27~19&      \nodata    &          ND      & \nodata          \nl
  L677-2    &19~19~52.8 & 11~30~46&      \nodata    &          ND      & \nodata          \nl
  L677-3    &19~20~07.5 & 11~15~30&      \nodata    &          ND      & \nodata          \nl
  L677-4    &19~20~22.8 & 11~16~57&      \nodata    &          ND      & \nodata          \nl
  L774-1    &19~20~44.9 & 23~19~21&   11.27$\pm$0.05&          \nodata & \nodata          \nl
  L774-2    &19~21~34.2 & 23~17~18&   ND            &          \nodata & \nodata          \nl
  L769-4    &19~22~47.0 & 22~56~36&      \nodata    &          ND      & \nodata          \nl
  L698      &19~23~37.5 & 12~59~48&      \nodata    &          ND      & \nodata          \nl
  L778-2    &19~25~22.0 & 23~39~45&   ND            &          ND      & \nodata          \nl
  L778-3    &19~25~54.8 & 23~37~14&   ND            &          ND      & \nodata          \nl
  L778-4    &19~26~02.6 & 23~51~12&      \nodata    &          ND      & \nodata          \nl
  L694-1    &19~38~06.5 & 10~43~33&      \nodata    &          ND      & \nodata          \nl
  L694-3    &19~38~54.0 & 10~43~42&   ND            &          \nodata & \nodata          \nl
  L758      &19~44~12.8 & 18~54~47&   ND            &          \nodata & \nodata          \nl
  L880      &20~11~49.4 & 40~03~59&   ND            &          \nodata & \nodata          \nl
  L917      &20~38~16.6 & 43~58~37&    3.10$\pm$0.03&          \nodata & \nodata          \nl
  L918      &20~39~02.5 & 43~54~57&    1.35$\pm$0.06&          \nodata & \nodata          \nl
  L1147**   &20~39~59.9 & 67~10~10&    2.78$\pm$0.02&          ND      & \nodata          \nl
  L896-2    &20~40~26.3 & 39~36~01&   -1.95$\pm$0.10&          \nodata & \nodata          \nl
  L896-1    &20~40~28.8 & 39~42~22&   ND            &          \nodata & \nodata          \nl
  L1155D    &20~43~17.9 & 67~26~12&    2.53$\pm$0.02&          ND      & \nodata          \nl
  L915      &20~43~23.0 & 42~58~52&   -0.80$\pm$0.10&          \nodata & \nodata          \nl
  L1082C-3  &20~47~59.4 & 60~00~36&   -2.90$\pm$0.01&          ND      & \nodata          \nl
  L981-2    &20~58~38.0 & 50~05~09&   -0.24$\pm$0.05&          \nodata & \nodata          \nl
  L971-1    &20~59~08.1 & 49~17~21&   -0.41$\pm$0.07&          \nodata & \nodata          \nl
  L971-2    &20~59~15.8 & 49~07~34&   -0.48$\pm$0.04&          \nodata & \nodata          \nl
  L1172A-2  &21~01~10.8 & 67~34~34&    2.82$\pm$0.03&          ND      & \nodata          \nl
  L1172A-3  &21~03~22.6 & 67~29~06&    2.89$\pm$0.02&          ND      & \nodata          \nl
  L1125-1   &21~13~07.8 & 61~32~26&    0.99$\pm$0.05&          \nodata & \nodata          \nl
  L944-1    &21~14~42.6 & 42~59~47&    ND           &          \nodata & \nodata          \nl
  L944-3    &21~16~59.7 & 43~12~57&    ND           &          \nodata & \nodata          \nl
  L952-1    &21~18~08.5 & 43~18~07&    ND           &          \nodata & \nodata          \nl
  L952-2    &21~18~25.0 & 43~19~59&    ND           &          \nodata & \nodata          \nl
  L953      &21~19~31.8 & 43~08~46&    ND           &          \nodata & \nodata          \nl
  L1021     &21~20~34.8 & 50~50~17&    2.78$\pm$0.10&          \nodata & \nodata          \nl
  L1014-1   &21~22~20.6 & 49~56~52&    4.02$\pm$0.07&          \nodata & \nodata          \nl
  L1014-2   &21~22~22.2 & 49~46~10&    ND           &          \nodata & \nodata          \nl
  L1099-1   &21~34~27.9 & 57~13~18&   -8.27$\pm$0.02&          ND      & \nodata          \nl
  L1111     &21~38~53.6 & 57~34~24&    ND           &          \nodata & \nodata          \nl
  L1103-1   &21~39~49.8 & 56~23~32&    ND           &          \nodata & \nodata          \nl
  L1103-2   &21~40~41.0 & 56~30~07&    4.44$\pm$0.03&          \nodata & \nodata          \nl
  L1113-2   &21~42~45.4 & 56~57~33&    ND           &          \nodata & \nodata          \nl
  L1113-1   &21~42~56.7 & 56~56~25&    ND           &          \nodata & \nodata          \nl
  L1139-1   &21~53~14.7 & 58~20~59&   -2.32$\pm$0.04&          \nodata & \nodata          \nl
  L1166     &22~03~46.7 & 59~19~07&    ND           &          \nodata & \nodata          \nl
  L1235B**  &22~15~04.1 & 73~07~30&   -4.19$\pm$0.02&          ND      &-4.00 $\pm$0.01   \nl
  L1195     &22~25~18.1 & 61~01~06&    ND           &          \nodata & \nodata          \nl
  L1185     &22~27~27.6 & 58~53~30&    ND           &          \nodata & \nodata          \nl
  L1229     &23~12~16.8 & 61~46~47&    ND           &          \nodata & \nodata          \nl
  L1263     &23~54~54.2 & 64~32~27&    ND           &          \nodata & \nodata          \nl
  L1271-1   &23~57~32.6 & 67~05~07&    ND           &          \nodata & \nodata          \nl
  L1271-3   &23~59~15.3 & 67~01~48&    ND           &          \nodata & \nodata          \nl
\enddata								        
\tablenotetext{a}{This lists velocity information for sources 
which have been observed in CS (2-1) and N$_2$H$^+$ (1-0), 
but detected in only one line or not detected in either line. 
The C$^{18}$O (1-0) line data which  have been complementarily 
obtained for several sources are also listed in this table.
`ND' and `\nodata' mean `No Detection' and `Not Observed', respectively.}

\tablenotetext{b}{Position of the source marked with `*' 
behind the source name is from Benson \& Myers (1989), `**' 
from the FCRAO map (Lee, Myers, \& Tafalla 1999), and none from LM99 catalogue.}
									         
\end{deluxetable}							         
        
\begin{deluxetable}{lrrrrrrrr}
\small
\tablecaption{ Line Parameters of 69 Starless Cores detected in both CS(2-1) and N$_2$H$^+$(1-0) lines}
\tablehead{
        \colhead{Source\tablenotemark{a}} &\colhead{RA (1950.0)}     & \colhead{DEC (1950.0)}          &
        \colhead{$V_{CS}$\tablenotemark{b}} &\colhead{$V_{N_2H^+}$\tablenotemark{c}} & \colhead{$V_{C^{18}O}$} & 
        \colhead{$\Delta V_{N_2H^+}$\tablenotemark{c}}   & 
        \colhead{$\tau_{N_2H^+}$\tablenotemark{c}}       & \colhead{$\delta V_{CS}$\tablenotemark{d}}   \nl
        \colhead{}                        & \colhead{$^h~~^m~~^s~$}  &\colhead{~~~~$^{\circ}~~~'~~~''$}&
        \colhead{(km~s$^{-1}$)}           & \colhead{(km~s$^{-1}$)}  & \colhead{(km~s$^{-1}$)}         & 
        \colhead{(km~s$^{-1}$)}           &\colhead{}                & \colhead{}
        }
\startdata
  L1333     &02~21~10.2 & 75~13~32&  3.83$\pm$0.02&  3.82$\pm$0.01& \nodata       & 0.38$\pm$0.01&  9.7$\pm$ 1.3 &  0.03$\pm$0.04 \nl
  L1355     &02~48~41.4 & 68~43~36& -3.91$\pm$0.01& -3.73$\pm$0.01& \nodata       & 0.26$\pm$0.03&  0.8$\pm$ 2.5 & -0.69$\pm$0.09 \nl
  B5-1      &03~44~26.4 & 32~46~45&  9.83$\pm$0.02&  9.93$\pm$0.06& \nodata       & 1.03$\pm$0.11& *0.1$\pm$ 0.7 & -0.10$\pm$0.06 \nl
  L1498*    &04~07~50.0 & 25~02~13&  7.66$\pm$0.01&  7.84$\pm$0.01&  7.95$\pm$0.02& 0.21$\pm$0.02&  5.7$\pm$ 2.6 & -0.86$\pm$0.08 \nl
  L1399-1   &04~20~00.5 & 55~01~00&  3.39$\pm$0.02&  3.26$\pm$0.03& \nodata       & 0.52$\pm$0.07&  1.0$\pm$ 0.1 &  0.26$\pm$0.08 \nl
  L1521B-2  &04~20~32.8 & 26~33~11&  6.46$\pm$0.02&  6.45$\pm$0.02& \nodata       & 0.19$\pm$0.07& *2.5$\pm$ 6.5 &  0.04$\pm$0.14 \nl
  L1399-2   &04~20~58.3 & 54~52~48&  3.49$\pm$0.02&  3.41$\pm$0.01& \nodata       & 0.45$\pm$0.03&  1.5$\pm$ 1.3 &  0.17$\pm$0.05 \nl
  L1399-3   &04~21~11.8 & 54~49~59&  3.37$\pm$0.03&  3.59$\pm$0.04& \nodata       & 0.62$\pm$0.07& 18.4$\pm$ 7.1 & -0.36$\pm$0.09 \nl
  B217-2    &04~25~04.5 & 26~14~16&  6.80$\pm$0.01&  6.86$\pm$0.01& \nodata       & 0.37$\pm$0.02&  4.2$\pm$ 1.2 & -0.18$\pm$0.04 \nl
  L1521F**  &04~25~35.0 & 26~45~00&  6.19$\pm$0.01&  6.51$\pm$0.00& *6.54$\pm$0.01& 0.35$\pm$0.01& 13.1$\pm$ 1.4 & -0.91$\pm$0.03 \nl
  L1521-2   &04~26~26.8 & 26~53~27&  6.62$\pm$0.01&  6.52$\pm$0.02& \nodata       & 0.25$\pm$0.04&*11.7$\pm$ 9.2 &  0.42$\pm$0.11 \nl
  L1400K**  &04~26~45.3 & 54~44~37&  3.04$\pm$0.01&  3.16$\pm$0.02& *3.12$\pm$0.01& 0.27$\pm$0.04&  5.0$\pm$ 3.6 & -0.43$\pm$0.08 \nl
  L1400A**  &04~26~50.6 & 54~46~08&  3.28$\pm$0.01&  3.31$\pm$0.01&  3.24$\pm$0.02& 0.23$\pm$0.03&  4.7$\pm$ 3.1 & -0.12$\pm$0.07 \nl
  L1524-4   &04~27~04.1 & 24~18~47&  6.11$\pm$0.01&  6.22$\pm$0.01& \nodata       & 0.32$\pm$0.01&  0.4$\pm$ 0.0 & -0.33$\pm$0.04 \nl
  L1551S-2  &04~28~03.5 & 18~09~09&  6.43$\pm$0.07&  6.69$\pm$0.02& \nodata       & 0.37$\pm$0.04&  0.5$\pm$ 0.0 & -0.71$\pm$0.21 \nl
  TMC2**    &04~29~46.9 & 24~18~54&  6.07$\pm$0.01&  6.23$\pm$0.01& *6.37$\pm$0.02& 0.30$\pm$0.02&  7.1$\pm$ 2.7 & -0.55$\pm$0.07 \nl
  B18-3     &04~31~04.6 & 24~03~32&  6.77$\pm$0.02&  6.96$\pm$0.03& \nodata       & 0.30$\pm$0.08& *0.1$\pm$21.2 & -0.67$\pm$0.22 \nl
  B18-5     &04~32~49.7 & 24~03~15&  6.29$\pm$0.02&  6.26$\pm$0.02& \nodata       & 0.37$\pm$0.04&  8.0$\pm$ 3.4 &  0.09$\pm$0.07 \nl
  CB22      &04~37~30.2 & 29~47~13&  5.92$\pm$0.02&  5.99$\pm$0.01& \nodata       & 0.31$\pm$0.03& *0.1$\pm$ 1.8 & -0.20$\pm$0.09 \nl
  TMC1**    &04~38~29.0 & 25~39~01&  5.74$\pm$0.01&  5.88$\pm$0.01& \nodata       & 0.43$\pm$0.02&  5.1$\pm$ 1.0 & -0.34$\pm$0.03 \nl
  L1507A-1  &04~39~29.0 & 29~38~07&  6.16$\pm$0.02&  6.22$\pm$0.01& *6.28$\pm$0.02& 0.22$\pm$0.02&  6.5$\pm$ 2.7 & -0.25$\pm$0.10 \nl
  CB23      &04~40~22.0 & 29~33~36&  5.91$\pm$0.01&  6.06$\pm$0.01& *6.01$\pm$0.07& 0.24$\pm$0.02&  5.9$\pm$ 2.6 & -0.67$\pm$0.09 \nl
  L1517A**  &04~51~58.7 & 30~28~04&  5.74$\pm$0.01&  5.79$\pm$0.02& *5.74$\pm$0.01& 0.21$\pm$0.06& *1.5$\pm$ 5.8 & -0.23$\pm$0.11 \nl
  L1517B*   &04~52~07.2 & 30~33~18&  5.80$\pm$0.01&  5.79$\pm$0.01& *5.85$\pm$0.02& 0.25$\pm$0.02&  1.3$\pm$ 1.7 &  0.02$\pm$0.04 \nl
  L1512-1*  &05~00~54.4 & 32~39~00&  7.16$\pm$0.01&  7.14$\pm$0.00& \nodata       & 0.15$\pm$0.01& 10.9$\pm$ 3.5 &  0.10$\pm$0.07 \nl
  L1544**   &05~01~12.5 & 25~06~40&  6.97$\pm$0.01&  7.14$\pm$0.01& \nodata       & 0.34$\pm$0.01& 11.2$\pm$ 2.2 &*-0.52$\pm$0.04 \nl
  L1552     &05~17~40.7 & 26~02~07&  7.58$\pm$0.01&  7.66$\pm$0.01& \nodata       & 0.22$\pm$0.02&  6.2$\pm$ 3.9 & -0.37$\pm$0.08 \nl
  L1582A    &05~29~15.0 & 12~28~57& 10.13$\pm$0.01& 10.23$\pm$0.03& \nodata       & 0.53$\pm$0.06& *0.7$\pm$ 3.1 & -0.19$\pm$0.06 \nl
  L1622A-2**&05~52~02.9 & 01~53~15&  0.99$\pm$0.01&  1.12$\pm$0.01& *1.20$\pm$0.02& 0.30$\pm$0.01&  8.2$\pm$ 1.5 & -0.45$\pm$0.05 \nl
  L1622A-1  &05~52~17.1 & 01~56~55&  1.12$\pm$0.01&  1.17$\pm$0.01& \nodata       & 0.33$\pm$0.01&  2.5$\pm$ 1.0 & -0.16$\pm$0.04 \nl
  CB45      &06~06~02.0 & 17~50~48&  0.42$\pm$0.02&  0.37$\pm$0.02& \nodata       & 0.56$\pm$0.04&  8.5$\pm$ 2.9 &  0.08$\pm$0.05 \nl
  L134A*    &15~50~58.1 &-04~26~36&  2.70$\pm$0.01&  2.70$\pm$0.01&  2.59$\pm$0.01& 0.37$\pm$0.02&  0.2$\pm$ 0.0 & -0.01$\pm$0.04 \nl
  L183B**   &15~51~30.0 &-02~42~51&  2.19$\pm$0.01&  2.48$\pm$0.00& *2.48$\pm$0.01& 0.24$\pm$0.00& 13.1$\pm$ 0.3 & -1.21$\pm$0.04 \nl
  L183*     &15~51~35.7 &-02~40~54&  2.29$\pm$0.03&  2.46$\pm$0.01& *2.44$\pm$0.01& 0.23$\pm$0.02&  4.9$\pm$ 2.4 & -0.73$\pm$0.14 \nl
  L1696A**  &16~25~30.0 &-24~11~42&  3.29$\pm$0.01&  3.39$\pm$0.01&  3.52$\pm$0.01& 0.23$\pm$0.02&  6.2$\pm$ 2.8 & -0.42$\pm$0.07 \nl
  L1696B**  &16~25~57.8 &-24~14~09&  3.21$\pm$0.02&  3.31$\pm$0.01& *3.28$\pm$0.01& 0.23$\pm$0.01& 15.3$\pm$ 1.7 & -0.42$\pm$0.10 \nl
  L1704-1   &16~27~49.8 &-23~35~49&  2.67$\pm$0.01&  2.69$\pm$0.01& \nodata       & 0.24$\pm$0.03&  0.5$\pm$ 3.6 & -0.09$\pm$0.07 \nl
  L1709B-2  &16~29~49.3 &-23~46~22&  2.60$\pm$0.01&  2.58$\pm$0.01& \nodata       & 0.34$\pm$0.02&  4.1$\pm$ 2.0 &  0.06$\pm$0.04 \nl
  L1689B**  &16~31~43.6 &-24~31~40&  3.36$\pm$0.01&  3.51$\pm$0.02& *3.60$\pm$0.01& 0.22$\pm$0.04& 11.6$\pm$ 8.2 & -0.64$\pm$0.14 \nl
  L1782-2   &16~39~36.4 &-19~38~00&  5.34$\pm$0.02&  5.40$\pm$0.02& \nodata       & 0.29$\pm$0.01&  1.0$\pm$ 0.0 & -0.20$\pm$0.08 \nl
  L158*     &16~44~37.7 &-13~54~03&  3.71$\pm$0.01&  3.88$\pm$0.01&  3.99$\pm$0.01& 0.21$\pm$0.02&  5.0$\pm$ 2.0 & -0.82$\pm$0.08 \nl
  L204C-2   &16~44~58.5 &-12~18~04&  4.16$\pm$0.01&  4.26$\pm$0.02& \nodata       & 0.45$\pm$0.05& *0.1$\pm$ 1.8 & -0.21$\pm$0.06 \nl
  L234E-1   &16~45~22.6 &-10~52~10&  2.93$\pm$0.02&  3.10$\pm$0.01& \nodata       & 0.34$\pm$0.02&  4.0$\pm$ 1.7 & -0.52$\pm$0.07 \nl
  L63**     &16~47~21.0 &-18~00~10&  5.92$\pm$0.02&  5.80$\pm$0.01& *5.81$\pm$0.01& 0.24$\pm$0.02&  5.5$\pm$ 2.8 &  0.53$\pm$0.10 \nl
  L462-1    &18~04~51.9 &-04~43~57&  7.88$\pm$0.02&  7.91$\pm$0.01& \nodata       & 0.43$\pm$0.04&  1.8$\pm$ 2.1 & -0.07$\pm$0.06 \nl
  L462-2    &18~04~56.9 &-04~40~57&  7.83$\pm$0.02&  7.92$\pm$0.01& \nodata       & 0.35$\pm$0.02&  4.9$\pm$ 1.7 & -0.28$\pm$0.06 \nl
  L492      &18~13~11.6 &-03~47~17&  7.54$\pm$0.01&  7.74$\pm$0.01& \nodata       & 0.52$\pm$0.02&  3.8$\pm$ 1.0 & -0.40$\pm$0.03 \nl
  L429-1    &18~14~28.2 &-08~14~40&  7.11$\pm$0.01&  6.87$\pm$0.01& \nodata       & 0.31$\pm$0.03&  5.3$\pm$ 3.0 & *0.79$\pm$0.09 \nl
  L648-1    &18~37~09.7 & 12~38~37&  4.04$\pm$0.02&  4.09$\pm$0.02& \nodata       & 0.29$\pm$0.03& *0.1$\pm$ 1.9 & -0.18$\pm$0.08 \nl
  L673-7    &19~19~14.3 & 11~15~30&  7.10$\pm$0.02&  7.14$\pm$0.01& \nodata       & 0.54$\pm$0.02&  2.9$\pm$ 1.1 & -0.08$\pm$0.04 \nl
  L694-2    &19~38~42.4 & 10~49~59&  9.39$\pm$0.01&  9.59$\pm$0.00& \nodata       & 0.27$\pm$0.00& 12.2$\pm$ 0.3 & -0.76$\pm$0.04 \nl
  L1041-2   &20~36~05.9 & 57~38~48& -1.80$\pm$0.02& -1.75$\pm$0.03& \nodata       & 0.39$\pm$0.05&  7.2$\pm$ 5.5 & -0.11$\pm$0.08 \nl
  L922-1    &20~36~46.5 & 44~21~24&  6.57$\pm$0.01&  6.63$\pm$0.01& \nodata       & 0.32$\pm$0.03&  3.8$\pm$ 2.7 & -0.19$\pm$0.06 \nl
  L922-2    &20~37~04.6 & 44~25~20&  6.45$\pm$0.01&  6.57$\pm$0.02& \nodata       & 0.35$\pm$0.04&  0.1$\pm$ 1.1 & -0.36$\pm$0.07 \nl
  L1049-1   &20~39~50.8 & 57~16~45&  0.25$\pm$0.02&  0.15$\pm$0.02& \nodata       & 0.49$\pm$0.03&  7.0$\pm$ 2.5 &  0.20$\pm$0.05 \nl
  L1049-2   &20~40~16.8 & 57~22~57&  0.20$\pm$0.04&  0.12$\pm$0.02& \nodata       & 0.41$\pm$0.05&  1.7$\pm$ 3.3 &  0.20$\pm$0.11 \nl
  L1148     &20~40~38.7 & 67~09~48&  2.53$\pm$0.02&  2.60$\pm$0.01& \nodata       & 0.29$\pm$0.03&  3.9$\pm$ 3.1 & -0.23$\pm$0.08 \nl
  L1155C-2**&20~42~52.1 & 67~39~36&  1.28$\pm$0.01&  1.39$\pm$0.01& \nodata       & 0.30$\pm$0.03&  5.9$\pm$ 2.9 & -0.36$\pm$0.06 \nl
  L1155C-1* &20~43~00.0 & 67~41~47&  2.44$\pm$0.01&  2.70$\pm$0.01&  2.83$\pm$0.03& 0.44$\pm$0.03&  5.0$\pm$ 2.0 & -0.60$\pm$0.05 \nl
  L1155E    &20~43~46.3 & 67~31~02&  2.40$\pm$0.03&  2.44$\pm$0.06& \nodata       & 0.62$\pm$0.11& *0.1$\pm$ 1.4 & -0.07$\pm$0.11 \nl
  L1082C-2  &20~48~38.9 & 60~04~25& -2.55$\pm$0.02& -2.52$\pm$0.01& \nodata       & 0.32$\pm$0.03&  3.0$\pm$ 2.4 & -0.08$\pm$0.07 \nl
  L981-1    &20~58~36.3 & 50~08~19& -0.21$\pm$0.02& -0.12$\pm$0.02& \nodata       & 0.53$\pm$0.02&  0.8$\pm$ 0.1 & -0.18$\pm$0.06 \nl
  L1063     &21~06~44.6 & 56~06~09& -0.78$\pm$0.02& -0.80$\pm$0.03& \nodata       & 0.39$\pm$0.06&  1.6$\pm$ 3.9 &  0.04$\pm$0.09 \nl
  L944-2    &21~15~54.0 & 43~05~41&  5.38$\pm$0.03&  5.37$\pm$0.04& \nodata       & 0.34$\pm$0.05&*30.0$\pm$22.8 &  0.04$\pm$0.14 \nl
  L1251A-2  &22~26~46.5 & 74~58~10& -3.98$\pm$0.02& -3.95$\pm$0.02& \nodata       & 0.51$\pm$0.03&  0.1$\pm$ 3.0 & -0.08$\pm$0.04 \nl
  L1197     &22~35~08.0 & 58~41~46& -3.25$\pm$0.02& -3.14$\pm$0.01& \nodata       & 0.28$\pm$0.03&  7.5$\pm$ 3.4 & -0.36$\pm$0.10 \nl
  L1251     &22~36~26.0 & 74~54~46& -3.63$\pm$0.02& -3.53$\pm$0.02& \nodata       & 0.56$\pm$0.04& *0.1$\pm$18.4 & -0.18$\pm$0.05 \nl
  L1234     &23~15~47.5 & 62~10~15& -5.20$\pm$0.02& -5.01$\pm$0.02& \nodata       & 0.41$\pm$0.04& *0.1$\pm$19.2 & -0.46$\pm$0.09 \nl
  CB246-2** &23~54~12.0 & 58~17~47& -0.71$\pm$0.02& -0.82$\pm$0.01&*-0.79$\pm$0.01& 0.29$\pm$0.03&  7.9$\pm$ 4.3 &  0.40$\pm$0.07 \nl
\enddata         

\tablenotetext{a} {Position references of sources marked with `*', `**', or
none behind the source name are the same as in Table 1.}

\tablenotetext{b}{The velocities of CS (2-1) spectra were obtained by a Gaussian fit. 
The given errors of velocities are 1 $\sigma$ values from the fit.}	             

\tablenotetext{c}{The velocities, FWHMs, and optical depths of N$_2$H$^+$ (1-0) spectra were obtained by the 
nonlinear Gaussian fit program in CLASS (Buisson et al. 1994). 
The $\tau_{N_2H^+}$ value with 
`*' is hard to be ensured  because of low S/N ratio ($\la 5$) of its spectrum. 
All designated errors  are 1 $\sigma$ values from the fitting.}

\tablenotetext{d}{The normalized velocity difference $\delta V_{CS}$ is defined as 
$ \delta V_{CS} = {{V_{CS}-V_{N_2H^+}} \over {\Delta V_{N_2H^+}}}$. 
The $1\sigma$ propagation errors are given 
at each $\delta V_{CS}$. Two sources (L1544 and L429-1) whose intensity ratio of blue to red
peak in CS spectra is close to 1.0 within error, are marked with `*' before 
the value of 
$\delta V_{CS}$. The CS peak velocities for these two were obtained by the Gaussian fit after masking 
a fainter peak between two components, though the difference between two is less than 1 $\sigma$.
We  do not use these $\delta V_{CS}$ values in our statistics.}
\end{deluxetable}

\begin{deluxetable}{lrcclrcc} 
\small
\tablecaption{Spectral Properties of Starless Cores}
\tablehead{
        \colhead{Source}   &\colhead{$\delta V_{CS}$\tablenotemark{a}}     & 
        \colhead{$\rm T_b/T_r$\tablenotemark{b} }          &
        \colhead{$\rm T_b/T_r$\tablenotemark{b} ~in} &
        \colhead{Source}   &\colhead{$\delta V_{CS}$\tablenotemark{a}}     &
        \colhead{$\rm T_b/T_r$\tablenotemark{b} }          &
        \colhead{$\rm T_b/T_r$\tablenotemark{b} ~in }  \nl
        \colhead{}         &\colhead{}                    & 
        \colhead{in CS}         &
        \colhead{$N_2H^+$ main component}            &
        \colhead{}         &\colhead{}                    &
        \colhead{in CS}          &
        \colhead{$N_2H^+$ main component}
        }
\startdata
  L1333     &      N&     \nodata       &         \nodata& L1696A   &       B&     \nodata      &         \nodata\nl
  L1355     &      B&  B (1.53$\pm$0.17)&         \nodata& L1696B   &       N&     \nodata      &         \nodata\nl
  B5-1      &      N&     \nodata       &         \nodata& L1704-1  &       N&     \nodata      &         \nodata\nl
  L1498     &      B&  B (2.03$\pm$0.17)&         \nodata& L1709B-2 &       N&     \nodata      &         \nodata\nl
  L1399-1   &      N&     \nodata       &         \nodata& L1689B   &       N& B (1.46$\pm$0.22)&         \nodata\nl
  L1521B-2  &      N&     \nodata       &         \nodata& L1782-2  &       N&     \nodata      &         \nodata\nl
  L1399-2   &      N&     \nodata       &         \nodata& L158     &       B& B (1.37$\pm$0.19)&         \nodata\nl
  L1521B    &\nodata&  N (0.91$\pm$0.19)&         \nodata& L204C    &\nodata & N (0.84$\pm$0.20)&         \nodata\nl
  L1399-3   &      N&     \nodata       &         \nodata& L204C-2  &       N&     \nodata      &         \nodata\nl
  B217-2    &      N&     \nodata       &         \nodata& L234E-1  &       B&     \nodata      &         \nodata\nl
  L1521F    &      B&  B (1.26$\pm$0.16)&B(1.13$\pm$0.12)& L63      &       R&     \nodata      &         \nodata\nl
  L1521-2   &      N&     \nodata       &         \nodata& L462-1   &       N&     \nodata      &         \nodata\nl
  L1400K    &      B&     \nodata       &         \nodata& L462-2   &       N&     \nodata      &         \nodata\nl
  L1400A    &      N&     \nodata       &         \nodata& L492     &       B&     \nodata      &N(1.15$\pm$0.16)\nl
  L1524-4   &      B&     \nodata       &         \nodata& L429-1   & \nodata& N (0.79$\pm$0.24)&N(1.43$\pm$0.45)\nl
  L1551S-2  &      N&     \nodata       &N(0.94$\pm$0.19)& L648-1   &       N&     \nodata      &         \nodata\nl
  L1445     &\nodata&  B (1.29$\pm$0.20)&         \nodata& L673-7   &       N&     \nodata      &         \nodata\nl
  TMC2      &      B&     \nodata       &         \nodata& L694-2   &       B& B (1.70$\pm$0.20)&         \nodata\nl
  B18-3     &      N&  B (1.24$\pm$0.20)&         \nodata& L1041-2  &       N&     \nodata      &         \nodata\nl
  B18-5     &      N&     \nodata       &         \nodata& L922-1   &       N&     \nodata      &         \nodata\nl
  CB22      &      N&     \nodata       &         \nodata& L922-2   &       B&     \nodata      &         \nodata\nl
  TMC1      &      B&     \nodata       &         \nodata& L1049-1  &       R&     \nodata      &         \nodata\nl
  L1507A-1  &      N&     \nodata       &         \nodata& L1049-2  &       N&     \nodata      &         \nodata\nl
  CB23      &      B&     \nodata       &         \nodata& L1148    &       N&     \nodata      &         \nodata\nl
  L1517C-1  &\nodata&  N (0.96$\pm$0.16)&         \nodata& L1155C-2 &       B&     \nodata      &         \nodata\nl
  L1517A    &      N&     \nodata       &         \nodata& L1155C-1 &       B& B (2.59$\pm$0.19)&         \nodata\nl
  L1517B    &      N&     \nodata       &         \nodata& L1155E   &       N&     \nodata      &         \nodata\nl
  L1512-1   &      N&     \nodata       &         \nodata& L1082C-2 &       N&     \nodata      &         \nodata\nl
  L1544     &\nodata&  N (1.13$\pm$0.20)&B(1.47$\pm$0.22)& L981-1   &       N&     \nodata      &         \nodata\nl
  L1552     &      N&     \nodata       &         \nodata& L1063    &       N&     \nodata      &         \nodata\nl
  L1582A    &      N&     \nodata       &         \nodata& L944-2   &       N&     \nodata      &         \nodata\nl
  L1622A-2  &      B&     \nodata       &         \nodata& L1251A-2 &       N&     \nodata      &         \nodata\nl
  L1622A-1  &      N&     \nodata       &         \nodata& L1197    &       N&     \nodata      &         \nodata\nl
  CB45      &      N&     \nodata       &         \nodata& L1251    &       N&     \nodata      &         \nodata\nl
  L134A     &      N&     \nodata       &         \nodata& L1234    &       B&     \nodata      &         \nodata\nl
  L183B     &      B&  B (1.46$\pm$0.22)&         \nodata& CB246-2  &       R& R (0.83$\pm$0.12)&         \nodata\nl
  L183      &      B&     \nodata       &         \nodata&          &        &                  &         \nodata\nl
\enddata			      
\tablenotetext{a}{`B' and `R' are given for sources with blue or red
asymmetry of $\vert \delta V_{CS}\vert \ge 5~\sigma_{\delta V_{CS}}$, 
and `N' for sources with neither asymmetry of 
$\vert \delta V_{CS}\vert \le 5~\sigma_{\delta V_{CS}}$}.

\tablenotetext{b}{The ratio of intensities of blue component ($\rm T_b$) and red component ($\rm T_r$)
from the Gaussian fit after masking the other fainter component.
`B', `N', or `R' are designated for the sources 
of $\rm T_b/T_r > 1+\sigma$, $\rm 1-\sigma \le T_b/T_r \le
1+\sigma$, or $\rm T_b/T_r < 1-\sigma$, respectively.} 

\end{deluxetable}